# Room temperature spin-orbit torque switching induced by a topological insulator

Jiahao Han[1], A. Richardella[2], Saima Siddiqui[1], Joseph Finley[1], N. Samarth[2] and Luqiao Liu[1*]

[1]Department of Electrical Engineering and Computer Science, Massachusetts Institute of Technology, Cambridge, Massachusetts 02139, USA

[2]Department of Physics, The Pennsylvania State University, University Park, Pennsylvania 16802, USA

*Email: luqiao@mit.edu

**Recent studies on the magneto-transport properties of topological insulators (TI)[1–7] have attracted great attention due to the rich spin-orbit physics and promising applications in spintronic devices. Particularly the strongly spin-moment coupled electronic states have been extensively pursued to realize efficient spin-orbit torque (SOT) switching. However, so far current-induced magnetic switching with TI has only been observed at cryogenic temperatures. It remains a controversial issue whether the topologically protected electronic states in TI could benefit spintronic applications at room temperature. In this work, we report full SOT switching in a TI/ferromagnet bilayer heterostructure with perpendicular magnetic anisotropy at room temperature. The low switching current density provides a definitive proof on the high SOT efficiency from TI. The effective spin Hall angle of TI is determined to be several times larger than commonly used heavy metals. Our results demonstrate the robustness of TI as an SOT switching material and provide a direct avenue towards applicable TI-based spintronic devices.**

Spin-orbit coupling has been extensively studied for the conversion between charge current and spin current[8]. When neighbored with a ferromagnet (FM), non-equilibrium spins induced by the spin-orbit coupling can exert torques onto magnetic moments (Fig.

1a), which remarkably manipulates the magnetic dynamics[9,10] and even switches the magnetization[11–13]. This current-induced switching is expected to lead to logic and memory device applications with high efficiency[8]. TI are a class of materials with spin-orbit coupling that is strong enough to invert the band structure and leads to complete spin-momentum locking in the surface states[1–4]. Recently it was shown that by utilizing the current-induced SOT from TI, one could switch the moments in magnetically doped TI heterostructures[6]. However, restricted by the low Curie temperature of the employed FM materials, the switching has only been realized at a few Kelvin. Room temperature SOT switching, which lies at the heart of applicative interests, remains to be demonstrated. On the other hand, several different techniques (including spin pumping[14–16], spin torque ferromagnetic resonance[4,5,17], second harmonic magnetometry[6,7], non-local spin valve or tunnel junction measurements[3,18–21], and spin Seebeck effect[22]) have been applied to extract the figure of merit for charge-spin conversion, the effective spin Hall angle $\alpha_{SH}$. The obtained results vary by orders of magnitude, which obscures fundamental understanding on the spin transport mechanism. Moreover, very controversial temperature dependencies of $\alpha_{SH}$ have been reported in different measurements. The question comes naturally whether the topologically protected electronic states would be robust enough for room temperature applications. Therefore, a direct, definitive experimental evidence to illustrate the SOT switching efficiency of TI at room temperature is highly desirable.

A major difficulty for realizing TI-based SOT switching at room temperature is to grow FM layers with appropriate magnetic anisotropy accompanied with TI. For commonly used FM materials such as Co and CoFeB, their magnetic anisotropy strongly relies on interfacial conditions, which usually favors an in-plane orientation when grown on TI. Comparatively, transition metal–rare earth ferrimagnetic alloys (e.g., $Co_xTb_{1-x}$) are more promising owing to their robust perpendicular magnetic anisotropy (PMA) in the bulk[23,24]. In CoTb alloy, the sublattices from Co and Tb elements are antiferromagnetically coupled. The magnetic properties such as anisotropy energy coefficient and magnetization can be tuned within a wide range through the engineering on chemical compositions. In our work, GaAs substrate/$Bi_2Se_3$ (7.4nm)/CoTb (4.6nm)/$SiN_x$ (3nm) multilayer stacks are deposited through a combination of molecular beam epitaxy[25] and magnetron sputtering. The atomic ratio of Tb element in CoTb alloy (0.23) and the film thickness (4.6 nm) are chosen to get an optimal PMA on the TI material $Bi_2Se_3$ (Fig. 1b). The films are then patterned to Hall bar structures for transport measurements (Fig. 1c). The anomalous Hall resistance ($R_H$) as a function of applied out-of-plane magnetic field ($H_z$) is plotted in Fig. 1d.

Current-induced switching is measured by sweeping a dc current. A bias magnetic field ($H_x$) of 1000 Oe is applied along the current direction to get deterministic switching polarities here. As plotted in Fig. 2a, the current switches the magnetization of the CoTb layer between up and down directions, corresponding to the Hall resistance of ± 0.6 Ω. The curve changes its polarity when the bias field is inversed (Fig. 2b), which is a typical characteristic of the SOT switching on PMA films[12]. The total resistance change (1.2 Ω) is comparable with the $R_H$–$H_z$ curve under the same in-plane bias field (1.4 Ω, see Supplementary Section 2), which demonstrates an almost complete magnetization switching by current. The critical current density for switching (~ 3 × 10$^6$ A/cm$^2$) is much smaller than the typical values obtained in heavy metal/FM structures with much thinner FM layer (usually in the magnitude of 10$^7$ A/cm$^2$, see refs. 11–13 and Figs. 4a,4c), suggesting the high SOT efficiency in $Bi_2Se_3$.

It is known that the spin-orbit coupled surface states, including both topologically trivial and non-trivial ones, are prone to the application of magnetic fields in the perpendicular-to-plane direction. Controversial results have been obtained previously with respect to the gap formation in the band structure when TI is covered with FM atoms[26,27]. Therefore, it remains a question whether TI could still exhibit efficient charge-spin conversion when it is adjacent to a strong FM with PMA, where the exchange-induced Zeeman interaction scale with the Curie temperature $T_c$ of the FM[28]. Our experiment provides a first-hand evidence on the robustness of the SOT in TI/strong FM bilayers ($T_c$ ~ 600K[23], in contrast to the dilute FM utilized earlier with $T_c$ ~ 10 K[6,7]).

We now turn to the quantitative calibration of the SOT efficiency in $Bi_2Se_3$/CoTb heterostructures. When a bias field is applied along the current direction, the damping-like term of SOT, which is most relevant to the magnetization switching[12], exerts an out-of-plane effective field inside the Néel type domain walls. This effective field is reflected by the horizontal shift of the $R_H$–$H_z$ hysteresis curve[10,29]. As shown in Fig. 3a, under a bias field of $H_x = +800$ Oe, the center of the $R_H$–$H_z$ curves shifts from zero to positive (negative) side in the presence of a positive (negative) current. We measure the $R_H$–$H_z$ hysteresis loops under a series of applied current and $H_x$, and plot the center of hysteresis loops ($H_z^{eff}$) as a function of the current density ($j_e$) (Fig. 3b). The slope defined as $\chi \equiv H_z^{eff}/j_e$ represents the damping-like SOT efficiency, the sign of which depends on the direction of $H_x$.

The dependence of the SOT efficiency on the bias field is summarized in Fig. 3c. $\chi$ grows linearly in magnitude for small $H_x$, and reaches saturation ($\chi^{sat}$) when $H_x$ is larger than 200 Oe. The evolution of $\chi$ as a function of $H_x$ can be explained by the chirality change of the domain walls in the CoTb layer. In our samples, the Dzyaloshinskii-Moriya interaction (DMI) appears either at the $Bi_2Se_3$/CoTb interface or in the bulk of the CoTb alloy, leading to the formation of the Néel domain walls with a certain chirality. As the bias field increases to overcome the DMI effective field ($H_{DMI}$), the domain walls change their chirality and start to move in directions that facilitate magnetic switching[10,24,29]. In this scenario, the saturation $\chi^{sat}$ represents the intrinsic SOT efficiency and the saturating bias field ($H_x^{eff} \approx 200$ Oe) can serve as an estimation of $H_{DMI}$ and the DMI energy (see Supplementary Section 3).

To compare the current-induced SOT efficiency of TI with other heavy metals, we calculate the effective spin Hall angle ($\alpha_{SH}$, defined as the ratio between the generated spin current density $\frac{2e}{\hbar}j_s$ and the average charge current density in TI $j_e$) of the $Bi_2Se_3$/CoTb heterostructure through $\chi^{sat}$ using[29]

$$\chi^{sat} = \frac{\pi}{2}\frac{\alpha_{SH}\hbar}{2\mu_0 M_s t} \quad (1)$$

where $\hbar$ is Planck's constant, $e$ is the electron charge, $\mu_0$ is the vacuum permeability, $M_s$ is the saturated magnetization, and $t$ is the thickness of the CoTb layer. The effective spin Hall angle is determined to be 0.16 ± 0.02. Noticeably, previously reported spin Hall angle of TI varies from 0.01 over 400[3–7,14–21]. Compared with other measurements, the method we use is straightforward and more relevant to the real application scenario, where the charge-spin conversion efficiency is quantified in the magnetic switching configurations. Besides the absolute values, very different trends have also been observed on the temperature dependence of the charge-spin conversion efficiency previously. While some experiments suggest the immunity of TI to temperature increase[5,16], many others indicate a very quick decay of the efficiency when the temperature is raised above the liquid helium temperature[3,14,17,18], which dims the prospect of TI for applicable switching devices. Our quantification of $\alpha_{SH}$ at room temperature, together with the results in Fig. 2, demonstrates the feasibility for room temperature applications.

To make a direct comparison on the SOT efficiencies between $Bi_2Se_3$ and other widely studied heavy metals, we fabricate additional control samples with Pt/CoTb/SiN$_x$ and Ta/CoTb/SiN$_x$ stacks. We first try the same CoTb thickness as in the $Bi_2Se_3$ sample. However, due to the reduced SOT efficiency in these metals, no current-induced switching can be detected. Alternatively, we grow thinner CoTb films (~ 2 nm) on Pt and Ta. Current-induced switching in these two samples is illustrated in Figs. 4a and 4c, while the SOT efficiency as a function of the bias field is summarized in Figs. 4b and 4d. First of all, from the current-induced switching polarity and SOT efficiency measurements, we note that the effective spin Hall angle of $Bi_2Se_3$ has the same sign with that of Pt but opposite with Ta, which is consistent with the previously reported results[3,4,16,21]. This also rules out the possibility that the spin-orbit coupling from the heavy element Tb plays the dominant role in the SOT switching, which will otherwise lead to same switching polarities in all of three samples. Secondly, the effective spin Hall angles of Pt and Ta are determined to be 0.017 ± 0.002 and − 0.031 ± 0.002 with Eq. (1), which is several times weaker than $Bi_2Se_3$ (Fig. 4e). Overall, the measured spin Hall angles from Pt and Ta are smaller than the numbers reported earlier (0.15 for Pt and − 0.12 for Ta), where similar techniques but different ferromagnetic electrodes (CoFeB and Co) have been utilized[29]. This may suggest that compared with conventional ferromagnetic electrodes, antiferromagnetically coupled CoTb layer tends have lower interfacial transparency[30] when it is in contact with spin-orbit materials. Therefore, the intrinsic effective spin Hall angle from $Bi_2Se_3$ could be even larger than the measured value in this work.

Despite the increased charge-spin conversion efficiencies in TI, it remained unclear whether the utilization of these materials will lead to energy advantages, as TI usually have much higher resistivity $\rho$ compared with other spin-orbit metals. To answer this question, we calculate the power consumption for switching FM electrodes in unit magnetic volume, which should be proportional to $\rho/\alpha_{SH}^2$, as the critical current density for switching scales with the inverse of $\alpha_{SH}$. As shown in Fig. 4f, $Bi_2Se_3$ still stands out as the favorable material when power consumption is used as the comparison metric. For further optimization the SOT efficiency, TI materials with more insulating bulk states will be beneficial, as the power dissipation in the bulk channels can be avoided. Finally, we note that in our current implementation of TI-induced switching, the shunting of current through the FM metal layer is significant. To fully exploit the efficiency of TI, magnetic semiconductors/insulators would be highly favorable. Potential candidates include rare earth iron garnet or barium ferrite with PMA[31,32].

In summary, spin-orbit torque magnetization switching induced by topological insulator $Bi_2Se_3$ is observed at room temperature. The effective spin Hall angle is precisely calibrated in $Bi_2Se_3$ and compared with heavy metals Pt and Ta. The large spin Hall angle suggests the robustness of charge-spin conversion in TI even at room temperature and capped with a strong ferromagnet with perpendicular magnetization. Accompanied with recent SOT studies in TI, our work adds to the technical significance of the emerging field of topological spintronics and potentially points to practicable innovations in TI-based switching devices.

**Methods**

**Sample deposition**: The $Bi_2Se_3$ sample was deposited on GaAs (111)A molecular beam epitaxy (See ref. 25 for detailed information). 20 nm of Se was used as the capping layer to prevent oxidation. The sample was then transferred through air into the magnetron sputtering chamber with base pressure of $4 \times 10^{-9}$ Torr. Se capping layer was removed by heating the sample to 250 °C for 1 hour. After the sample was cooled to room temperature, the CoTb alloy layers were deposited on $Bi_2Se_3$, followed by an insulating $Si_3N_4$ layer (3 nm) to avoid oxidation.

**Device fabrication and electrical measurements:** The multilayer films were patterned to Hall devices with the width of 4 ~ 20 μm using photo lithography and argon ion milling. The electrical contacts consisting of Ta(10 nm)/Ru (80 nm) were deposited by magnetron sputtering after the $SiN_x$ was etched away in the contact areas. The resistivity of the $Bi_2Se_3$ and CoTb layers were determined using four-point measurements to be 1060 and 97 μΩcm, which is also consistent with the measured total resistance of $Bi_2Se_3$/CoTb/$SiN_x$ multilayer samples. The current density in $Bi_2Se_3$ was calculated using a parallel circuit model with these resistivity values. The resistivity of Ta and Pt was determined to be 193 and 23 μΩcm. For current-induced switching and SOT efficiency measurements, more than five devices have been tested and the results are similar to each other. The reported data are from the same device.

**Acknowledgements**

This research was partially supported by NSF through the Massachusetts Institute of Technology Materials Research Science and Engineering Center DMR - 1419807. The $Bi_2Se_3$ sample for this publication was provided by The Pennsylvania State University Two-Dimensional Crystal Consortium – Materials Innovation Platform (2DCC-MIP) which is supported by NSF cooperative agreement DMR-1539916.

**Author contributions**

J.H. fabricated the samples and performed the electrical measurements with the help of S.S. and J.F.. A.R. and N.S. deposited the $Bi_2Se_3$ samples. J.H. analyzed the data. J.H. and L.L. wrote the manuscript. All authors discussed the results and commented on the manuscript.

**Competing financial interests**

The authors declare no competing financial interests.

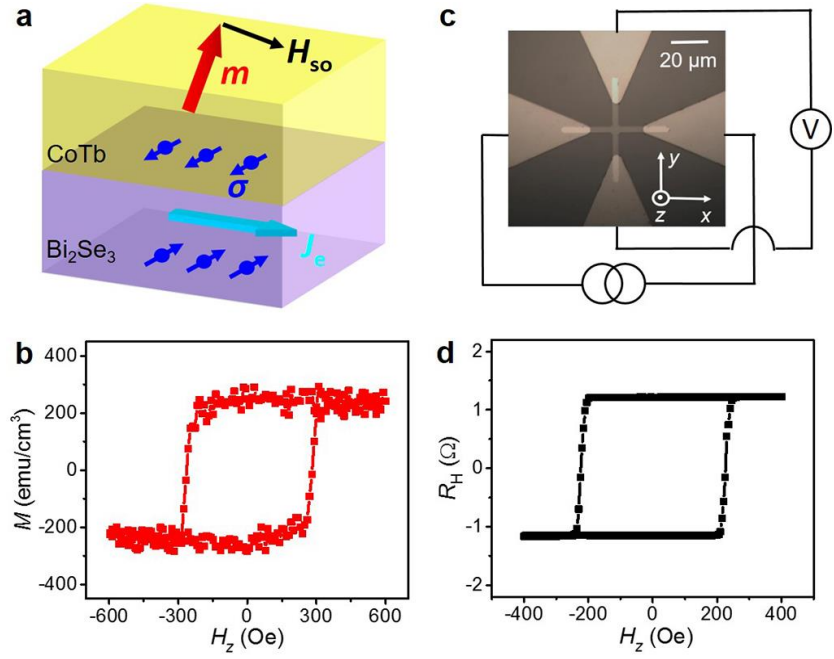

**Figure 1. Structure and magnetic properties of the Bi₂Se₃/CoTb sample. a**, Schematic of SOT in Bi$_2$Se$_3$/CoTb heterostructure. The charge current ($j_e$) generates spin accumulation ($\sigma$) at the interface due to the spin-orbit coupling, and exerts a SOT to the magnetic moments (*m*). The SOT is described by an effective field $H_{SO}$. **b**, Hysteresis loop of the out-of-plane magnetization measured by vibrating sample magnetometry (VSM). **c**, Image of the Hall device with the illustration of the electrical measurement setup. The current is applied along the *x* axis and the Hall voltage is detected in the *y* direction. The width of the Hall bar is 4 μm. **d**, Anomalous Hall resistance vs out-of-plane magnetic field.

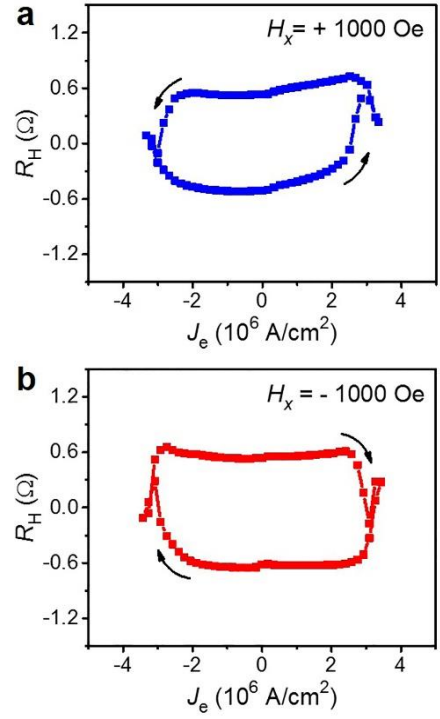

**Figure 2. Room temperature SOT switching in Bi$_2$Se$_3$/CoTb.** Hall resistance is measured when sweeping a dc current under a bias field of 1000 Oe along **a**, positive and **b**, negative *x* axis. $j_e$ is the average current density inside the Bi$_2$Se$_3$ layer.

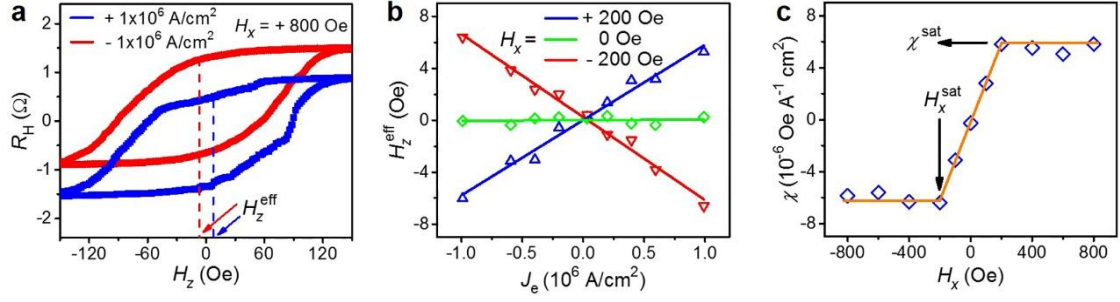

**Figure 3. Calibration of the SOT efficiency. a**, Hall resistance vs applied perpendicular field under positive and negative dc currents with the density of $1.0 \times 10^6$ A/cm$^2$ through Bi$_2$Se$_3$ (corresponding to the total current of 3 mA) and a bias field $H_x$ = + 800 Oe. The center shift corresponds to the SOT-induced effective field ($H_z^{eff}$). **b**, $H_z^{eff}$ as a function of applied current under bias fields $H_x$ = 0, ± 200 Oe. **c**, SOT efficiency $\chi$ vs $H_x$. $\chi$ saturates at the field $H_x^{sat}$.

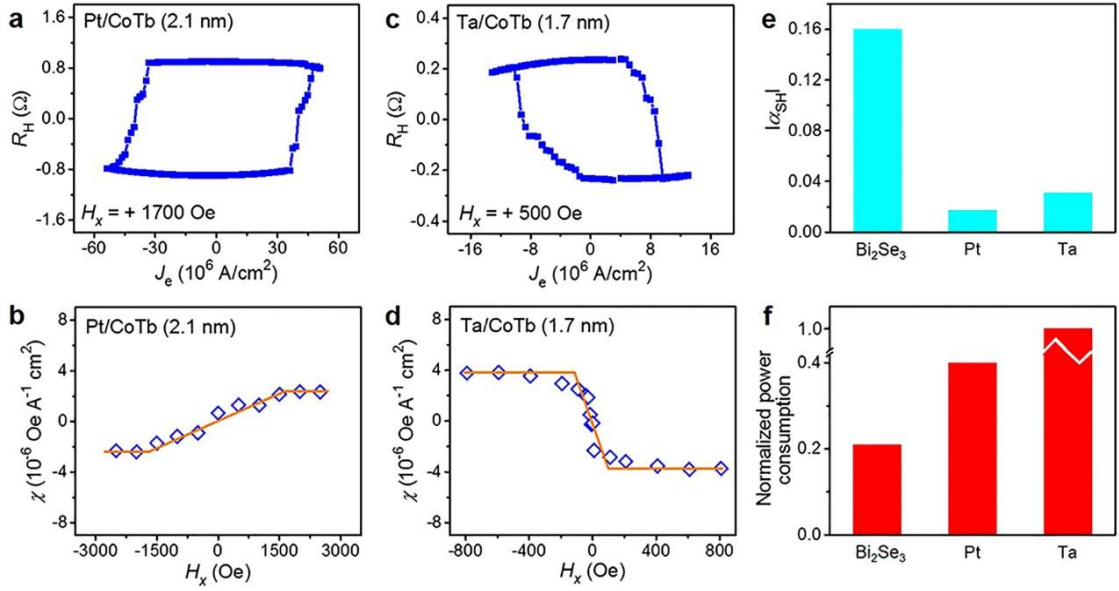

**Figure 4. Comparative measurements in CoTb grown on different spin-orbit materials. a,c**, Current-induced switching in Pt/CoTb and Ta/CoTb samples under positive bias fields. **b,d**, SOT efficiency $\chi$ vs bias field $H_x$ in Pt/CoTb and Ta/CoTb samples. Note that compared with the $Bi_2Se_3$/CoTb sample, CoTb layers with reduced thickness have been used for Pt and Ta samples, to allow for measurable current-induced switching. **e**, Absolute values of the effective spin Hall angles of $Bi_2Se_3$, Pt, and Ta measured by our experiments. **f**, Normalized power consumption (with Ta set to be unity) for switching FM electrodes in unit magnetic volume using $Bi_2Se_3$, Pt, and Ta.